Robust dual-field optimization of scanned ion beams against range and setup uncertainties

Taku Inaniwa,[a)] Nobuyuki Kanematsu, Takuji Furukawa and Koji Noda

*Medical Physics Research Group, Research Center for Charged Particle Therapy, National Institute of Radiological Sciences, 4-9-1 Anagawa, Inage-ku, Chiba 263-8555, Japan*

Email: taku@nirs.go.jp

Tel: +81-43-206-3170

Fax: +81-43-251-1840

Abstract

A 'dual-field' strategy is often used for tumors with highly complex shapes and/or with large volumes exceeding available field-size in both passive and scanning irradiations with ion beams. Range and setup uncertainties can cause hot and cold doses at the field junction within the target. Such uncertainties will also cause cold doses in the peripheral region of the target. We have developed an algorithm to reduce the sensitivity of the dual-field plan to these uncertainties in scanning irradiations. This algorithm is composed of the following two steps: 1) generating the expanded target volume, and 2) solving the inverse problem where the terms suppressing the dose gradient of individual fields are added into the objective function. The validity of this algorithm is demonstrated through the simulation studies for three extreme cases of two fields with unidirectional, opposing and orthogonal geometries. With the proposed algorithm, we can obtain a more robust plan to minimize the effects of range and setup uncertainties than the conventional plan. Compared to that for the conventional plan, the optimization time for the robust plan increased by a factor of approximately three.

1. Introduction

In the past decade, there has been growing interest in cancer therapy using heavy charged particles such as protons or heavier ions like carbon. This can be attributed to the good physical selectivity of beams composed of these particles, resulting in an inverted dose profile with a sharp longitudinal dose fall-off at the end of the particle range and a small lateral scattering. In addition, in the case of carbon, the increased biological effectiveness around the Bragg peak enhances its usefulness for radiotherapy. With these advantageous characteristics, charged particle therapy offers a high degree of dose conformity to a tumor while preventing undesired exposures of the surrounding normal tissues. For tumors with complex shapes like L-shaped targets in the head and neck, a 'dual-field' strategy is often used to avoid the critical structures located close to the tumor (Bussiere and Adams 2003). In this strategy, two fields are combined such that the first field covers only a part of the target, avoiding a nearby critical structure, and the second field covers the remaining portion of the target. The 'dual-field' strategy is also used for large volumes exceeding the available field-size, which is limited by the cross-sectional sizes of the beam shaping/controlling devices on the beam line, while that in the beam direction is limited by the maximum beam energy. In this case also, the first field is designed to partially cover the target, and the second field is used to cover the portion of the target that is not covered by the first field. However, the 'dual-field' strategy is sensitive to range and setup uncertainties in both passive (Paganetti et al 2008) and pencil beam scanning methods, i.e., intensity modulated proton/ion therapy (IMPT/IMIT) (Weber et al 2005, Rutz et al 2008, Albertini et al 2008, Lomax 2008). Range and setup errors cause hot and cold doses within the target and cold doses in the peripheral region of the target volume. To reduce these undesired effects, several technical approaches have been applied in passive irradiation methods (Hug et al 2000, Li et al 2007). In IMPT/IMIT with pencil beam scanning, there usually exist many different solutions to the inverse problem that will realize dosimetrically equivalent plans (Lomax et al 2004). This redundancy of solutions can be used to reduce the sensitivity of treatment plans if these uncertainties are accounted for in the optimization. Recently, several such approaches were proposed in IMPT (Unkelbach et al 2007, Pflugfelder et al 2008, Unkelbach et al 2009), and these methods greatly reduced the sensitivity to these uncertainties of the resulting treatment plans. However, in these approaches many possible scenarios must be considered, and for all these scenarios the resultant dose distributions must be calculated in each iterative optimization process, leading to computation time as long as several hours (Unkelbach et al 2009).

In this paper, we propose a simple and fast algorithm for reducing the sensitivity of the dual-field plan to these uncertainties. In our algorithm, only a nominal scenario is considered in each iterative optimization process by adding the terms for the sensitivity to the uncertainties into the objective function to lead to a robust solution directly. The proposed algorithm is fully integrated into the research version of treatment planning software developed for carbon-ion scanning (Inaniwa et al

2008). The validity of the proposed algorithm is demonstrated through the simulation studies for three extreme cases of two fields with unidirectional, opposing and orthogonal geometries.

In IMPT/IMIT, the non-uniform dose distributions are delivered from several directions, and the desired target coverage and the sparing of organs at risk (OARs) are obtained after superposing the dose contributions from all fields. For simplicity, in this paper, we do not consider the OARs, treating only target coverage in multi-field irradiations with scanned ion beams. Hence, we refer to the proposed algorithm as 'a robust approach for dual-field optimizations' rather than 'a robust approach for IMPT/IMIT'. Furthermore, for simplicity, only the physical dose distribution is considered in this paper.

2. Preliminary Calculations

2.1 Conventional optimization

A pencil beam algorithm is used for dose calculation. In the inverse planning, the dose-based objective function, $f(\mathbf{w})$, is minimized by an iterative process to determine the best particle numbers (beam weight: $w$) for each pencil beam. Here, $\mathbf{w}$ is the matrix notation of the beam weights for all pencil beams. The objective function is defined as

$$f(\mathbf{w}) = \sum_{i \in \mathrm{T}} \left( Q_\mathrm{P}^\mathrm{o} \, \mathrm{H}'\left[ \sum_{l=1}^{N_{field}} D_{i,l}(\mathbf{w}_l) - D_\mathrm{P}^{\max} \right]^2 + Q_\mathrm{P}^\mathrm{u} \, \mathrm{H}'\left[ D_\mathrm{P}^{\min} - \sum_{l=1}^{N_{field}} D_{i,l}(\mathbf{w}_l) \right]^2 \right), \quad (1)$$

where $N_{field}$ is the number of incident fields, and $D_{i,l}(\mathbf{w}_l)$ is the dose delivered to a position $i$ from an incident field $l$ with the beam weight $\mathbf{w}_l$. The total beam weight $\mathbf{w}$ has the following relationship to the matrix $\mathbf{w}_l$ for each incident field $l$:

$$\mathbf{w} = \sum_{l=1}^{N_{field}} \mathbf{w}_l. \quad (2)$$

$D_\mathrm{P}^{\max}$, $D_\mathrm{P}^{\min}$, $Q_\mathrm{P}^\mathrm{o}$, $Q_\mathrm{P}^\mathrm{u}$ are the maximum and minimum doses applied to the target T and the penalty coefficients for over- and under-dosage specified for the target, respectively. $\mathrm{H}'[r]$ is described as $\mathrm{H}'[r] = r\,\mathrm{H}[r]$ with the heaviside step function, $\mathrm{H}[r]$, defined so as to take the value of 1 only if $r$ is greater than zero; otherwise, it takes the value of 0. In this paper, the Bragg peak position of the pencil beam is referred to as 'spot'.

2.2 Pencil beam model

The details of the beam model used in this study were described in Inaniwa et al (2008). Therefore, the explanation of the model is kept to a minimum here. The $x$ and $y$ coordinates denote the lateral and orthogonal directions, respectively, and the $z$ coordinate denotes the direction parallel to the beam axis. The dose distribution delivered by the $j$-th pencil beam, $d_j(x, y, z)$, is split into three components, two components in transverse directions, $d_{x:j}(x,z)$ and $d_{y:j}(y,z)$, and one, $d_{z:j}(z)$, parallel to the beam direction, and represented as follows:

$$d_j(x,y,z) = d_{x:j}(x,z) d_{y:j}(y,z) d_{z:j}(z). \tag{3}$$

Here, $d_{x:j}(x,z)$ and $d_{y:j}(y,z)$ are the normalized Gaussian functions with standard deviations $\sigma_{x:j}(z)$ and $\sigma_{y:j}(z)$ representing the beam spread at a depth $z$ described by

$$d_{x:j}(x,z) = \frac{1}{\sqrt{2\pi}\sigma_{x:j}(z)} \exp\left(-\frac{(x-x_j)^2}{2\sigma_{x:j}(z)^2}\right) \tag{4}$$

and

$$d_{y:j}(x,z) = \frac{1}{\sqrt{2\pi}\sigma_{y:j}(z)} \exp\left(-\frac{(y-y_j)^2}{2\sigma_{y:j}(z)^2}\right), \tag{5}$$

respectively. On the other hand, $d_{z:j}(z)$ is the planner-integrated dose at a depth of $z$.

3. Robust Optimization Algorithm

3.1 Step1: Expansion of the target

As a first step of the robust optimization method, the target volume is expanded to account for the cold doses in the peripheral region of the target volume due to the range and setup errors. The magnitude of the error is non-isotropic and field specific. Hence, we expand the target volume to a volume that encompasses the target plus margins against non-isotropic errors for each incident field direction. The expanded target volume is referred to as the 'field-specific target volume (FTV)' in this paper. These errors can be split into a component parallel and a component perpendicular to the beam axis. The range uncertainties are the component parallel to the beam axis. There are multiple sources of range uncertainties, e.g., CT artifacts, weight gain or weight loss of a patient and conversion from Hounsfield units (HU) to stopping powers. In this study, we assume that the range uncertainties are proportional to the water equivalent depth, and they amount to ±5.0% as an extreme case. On the other hand, the setup errors comprise components both parallel and perpendicular to the beam axis. Shifts parallel to the beam axis cause an increase/decrease of the air gap in front of the patient, but this alters the resulting dose distribution only minimally. Thus, it is sufficient to consider only shifts perpendicular to each treatment beam to account for setup uncertainty (Pflugfelder et al 2008). We assumed setup uncertainties of ±5 mm in the *x*- and *y*-directions, as assumed in Lomax (2008). Spots are placed within the FTV in each incident field. On the other hand, in dose optimizations, a volume including the FTVs in all incident fields is considered as a dosimetric target volume (DTV).

3.2 Step2: Inverse Planning
3.2.1 Objective function

The authors of previous studies pointed out that steep longitudinal dose gradients make treatment plans sensitive to range errors, and steep lateral dose gradients make plans sensitive to setup errors (Pflugfelder et al 2008, Unkelbach et al 2009, Lomax 2008). Accordingly, as a robust optimization, we directly add the terms suppressing the in-field dose gradient within the target volume to the objective function (1), as follows:

$$f(\mathbf{w}) = \sum_{i \in T} \left( Q_P^o \, H' \left[ \sum_{l=1}^{N_{field}} D_{i,l}(\mathbf{w}_l) - D_P^{max} \right]^2 + Q_P^u \, H' \left[ D_P^{min} - \sum_{l=1}^{N_{field}} D_{i,l}(\mathbf{w}_l) \right]^2 \right)$$
$$+ \sum_{i \in T} \left( \sum_{l=1}^{N_{field}} \left( Q_x \left( \frac{\partial D_{i,l}(\mathbf{w}_l)}{\partial x} \right)^2 + Q_y \left( \frac{\partial D_{i,l}(\mathbf{w}_l)}{\partial y} \right)^2 + Q_z \left( \frac{\partial D_{i,l}(\mathbf{w}_l)}{\partial z} \right)^2 \right) \right)$$
, (6)

where $Q_x$, $Q_y$, $Q_z$ are the penalty coefficients for dose gradients in the two scanning directions and beam direction, respectively. $D_{i,l}(\mathbf{w}_l)$ can be described by the superposition of the dose of individual Bragg peaks, $d_j(x, y, z)$, according to their weights, $w_j$, as follows:

$$D_{i,l}(\mathbf{w}_l) = \sum_{j=1}^{N_l} d_j(x_i, y_i, z_i) w_j \equiv \sum_{j=1}^{N_l} d_{ij} w_j \quad , \qquad (7)$$

where $N_l$ is the number of spots in the field *l*.
By using equation (7), the objective function (6) can be transformed as follows;

$$f(\mathbf{w}) = \sum_{i \in \mathrm{T}} \left( Q_{\mathrm{P}}^{\mathrm{o}} \, \mathrm{H}' \left[ \sum_{l=1}^{N_{field}} \left( \sum_{j=1}^{N_l} d_{ij} w_j \right) - D_{\mathrm{P}}^{\max} \right]^2 + Q_{\mathrm{P}}^{\mathrm{u}} \, \mathrm{H}' \left[ D_{\mathrm{P}}^{\min} - \sum_{l=1}^{N_{field}} \left( \sum_{j=1}^{N_l} d_{ij} w_j \right) \right]^2 \right)$$

$$+ \sum_{i \in \mathrm{T}} \left( \sum_{l=1}^{N_{field}} \left( Q_x \left( \frac{\partial}{\partial x} \left( \sum_{j=1}^{N_l} d_{ij} w_j \right) \right)^2 + Q_y \left( \frac{\partial}{\partial y} \left( \sum_{j=1}^{N_l} d_{ij} w_j \right) \right)^2 + Q_z \left( \frac{\partial}{\partial z} \left( \sum_{j=1}^{N_l} d_{ij} w_j \right) \right)^2 \right) \right)$$

$$= \sum_{i \in \mathrm{T}} \left( \begin{array}{l} Q_{\mathrm{P}}^{\mathrm{o}} \, \mathrm{H}' \left[ \sum_{l=1}^{N_{field}} \left( \sum_{j=1}^{N_l} d_{ij} w_j \right) - D_{\mathrm{P}}^{\max} \right]^2 + Q_{\mathrm{P}}^{\mathrm{u}} \, \mathrm{H}' \left[ D_{\mathrm{P}}^{\min} - \sum_{l=1}^{N_{field}} \left( \sum_{j=1}^{N_l} d_{ij} w_j \right) \right]^2 + \\ Q_x \left( \sum_{l=1}^{N_{field}} \left( \sum_{j=1}^{N_l} \left( \frac{\partial d_{ij}}{\partial x} \right) w_j \right) \right)^2 + Q_y \left( \sum_{l=1}^{N_{field}} \left( \sum_{j=1}^{N_l} \left( \frac{\partial d_{ij}}{\partial y} \right) w_j \right) \right)^2 + Q_z \left( \sum_{l=1}^{N_{field}} \left( \sum_{j=1}^{N_l} \left( \frac{\partial d_{ij}}{\partial z} \right) w_j \right) \right)^2 \end{array} \right)$$

(8)

The additional dose gradient suppressing terms can be derived by the convolution of the gradients $(\partial d_{ij}/\partial x)$, $(\partial d_{ij}/\partial y)$ and $(\partial d_{ij}/\partial z)$ with respect to its weight, $w_j$, as $\sum_{l=1}^{N_{field}} \left( \sum_{j=1}^{N_l} \left( \frac{\partial d_{ij}}{\partial x} \right) w_j \right)$, $\sum_{l=1}^{N_{field}} \left( \sum_{j=1}^{N_l} \left( \frac{\partial d_{ij}}{\partial y} \right) w_j \right)$ and $\sum_{l=1}^{N_{field}} \left( \sum_{j=1}^{N_l} \left( \frac{\partial d_{ij}}{\partial z} \right) w_j \right)$, respectively. A similar convolution, i.e., the convolution of dose $d_{ij}$ with respect to its weight $w_j$, $\sum_{l=1}^{N_{field}} \left( \sum_{j=1}^{N_l} d_{ij} w_j \right)$, must be applied to derive the original terms.

To solve the inverse problem using the quasi-Newton method, the objective function's first derivative, $\nabla f(\mathbf{w})$, has to be calculated. The $j$-th component of $\nabla f(\mathbf{w})$ is derived as

$$[\nabla f(\mathbf{w})]_j = \frac{\partial f(\mathbf{w})}{\partial w_j}$$

$$= 2 \sum_{i \in \mathrm{T}} \left( \begin{array}{l} Q_{\mathrm{P}}^{\mathrm{o}} \, \mathrm{H}' \left[ \sum_{l=1}^{N_{field}} \left( \sum_{j'=1}^{N_l} d_{ij'} w_{j'} \right) - D_{\mathrm{P}}^{\max} \right] d_{ij} - Q_{\mathrm{P}}^{\mathrm{u}} \, \mathrm{H}' \left[ D_{\mathrm{P}}^{\min} - \sum_{l=1}^{N_{field}} \left( \sum_{j'=N_{1l}}^{N_l} d_{ij'} w_{j'} \right) \right] d_{ij} + \\ Q_x \left( \sum_{l=1}^{N_{field}} \left( \sum_{j'=1}^{N_l} \left( \frac{\partial d_{ij'}}{\partial x} \right) w_{j'} \right) \right) \left( \frac{\partial d_{ij}}{\partial x} \right) + Q_y \left( \sum_{l=1}^{N_{field}} \left( \sum_{j=1}^{N_l} \left( \frac{\partial d_{ij'}}{\partial y} \right) w_{j'} \right) \right) \left( \frac{\partial d_{ij}}{\partial y} \right) + Q_z \left( \sum_{l=1}^{N_{field}} \left( \sum_{j=N_1}^{N_l} \left( \frac{\partial d_{ij'}}{\partial z} \right) w_{j'} \right) \right) \left( \frac{\partial d_{ij}}{\partial z} \right) \end{array} \right)$$

. (9)

The quantities originating from the additional terms can be derived with the same functional form as that from the original terms.

### 3.2.2 Gradients of $d_{ij}$

As described in equations (8) and (9), the gradients of $d_{ij}$ in the $x$-, $y$- and $z$-directions, $(\partial d_{ij}/\partial x)$, $(\partial d_{ij}/\partial y)$ and $(\partial d_{ij}/\partial z)$, respectively, have to be calculated for derivations of $f(\mathbf{w})$ and $\nabla f(\mathbf{w})$. In this subsection, we derive the quantities $(\partial d_{ij}/\partial x)$, $(\partial d_{ij}/\partial y)$ and $(\partial d_{ij}/\partial z)$.

First, we derive $(\partial d_{ij}/\partial x)$ as follows:

$$\begin{aligned}
\frac{\partial d_{ij}}{\partial x} &= \frac{\partial}{\partial x}\left(d_j(x_i, y_i, z_i)\right) \\
&= \frac{\partial}{\partial x}\left(d_{x:j}(x_i, z_i) d_{y:j}(y_i, z_i) d_{z:j}(z_i)\right) \\
&\equiv \frac{\partial}{\partial x}\left(d_{x:ij} d_{y:ij} d_{z:ij}\right) \\
&= \frac{\partial}{\partial x}(d_{x:ij}) d_{y:ij} d_{z:ij} \\
&= \frac{\partial}{\partial x}\left(\frac{1}{\sqrt{2\pi}\sigma_{x:j}(z_i)} \exp\left(-\frac{(x_i - x_j)^2}{2\sigma_{x:j}(z_i)^2}\right)\right) d_{y:ij} d_{z:ij} \\
&= -\frac{(x_i - x_j)}{\sqrt{2\pi}\sigma_{x:j}(z_i)^3} \exp\left(-\frac{(x_i - x_j)^2}{2\sigma_{x:j}(z_i)^2}\right) d_{y:ij} d_{z:ij}
\end{aligned} \quad (10)$$

With the same transformations, we can derive $(\partial d_{ij}/\partial y)$ as follows:

$$\frac{\partial d_{ij}}{\partial y} = \frac{\partial}{\partial y}(d_{y:ij}) d_{x:ij} d_{z:ij} = -\frac{(y_i - y_j)}{\sqrt{2\pi}\sigma_{y:j}(z_i)^3} \exp\left(-\frac{(y_i - y_j)^2}{2\sigma_{y:j}(z_i)^2}\right) d_{x:ij} d_{z:ij}. \quad (11)$$

The functions

$$-\frac{(x - x_j)}{\sqrt{2\pi}\sigma_{x:j}(z)^3} \exp\left(-\frac{(x - x_j)^2}{2\sigma_{x:j}(z)^2}\right) \quad \text{and} \quad -\frac{(y - y_j)}{\sqrt{2\pi}\sigma_{y:j}(z)^3} \exp\left(-\frac{(y - y_j)^2}{2\sigma_{y:j}(z)^2}\right)$$

are the derivative of the normalized Gaussian functions with respect to the $x$- and $y$-directions, respectively.

On the other hand, the gradient of $d_{ij}$ in the beam-direction, $(\partial d_{ij}/\partial z)$, can be derived as follows:

$$\begin{aligned}
\frac{\partial d_{ij}}{\partial z} &= \frac{\partial}{\partial z}\left(d_{x:ij}d_{y:ij}d_{z:ij}\right) \\
&= d_{y:ij}d_{z:ij}\frac{\partial}{\partial z}\left(d_{x:ij}\right) + d_{x:ij}d_{z:ij}\frac{\partial}{\partial z}\left(d_{y:ij}\right) + d_{x:ij}d_{y:ij}\frac{\partial}{\partial z}\left(d_{z:ij}\right) \\
&= \frac{1}{\sigma_{x:j}(z_i)}\left(\frac{(x_i - x_j)^2}{\sigma_{x:j}(z_i)^2} - 1\right)d_{x:ij}d_{y:ij}d_{z:ij}\frac{\partial \sigma_{x:j}(z_i)}{\partial z} \\
&\quad + \frac{1}{\sigma_{y:j}(z_i)}\left(\frac{(y_i - y_j)^2}{\sigma_{y:j}(z_i)^2} - 1\right)d_{x:ij}d_{y:ij}d_{z:ij}\frac{\partial \sigma_{y:j}(z_i)}{\partial z} \\
&\quad + d_{x:ij}d_{y:ij}\frac{\partial d_{z:ij}}{\partial z}
\end{aligned} \qquad (12)$$

3.2.3   Implementation

In equations (8) and (9), the convolutions $\sum_{l=1}^{N_{field}}\left(\sum_{j=1}^{N_l}\left(\frac{\partial d_{ij}}{\partial x}\right)w_j\right)$ and $\sum_{l=1}^{N_{field}}\left(\sum_{j=1}^{N_l}\left(\frac{\partial d_{ij}}{\partial y}\right)w_j\right)$ of the additional terms can be derived by just replacing the normalized Gaussian filters $d_{x:ij}$ and $d_{y:ij}$ in $\sum_{l=1}^{N_{field}}\left(\sum_{j=1}^{N_l}d_{ij}w_j\right)$ of the original terms with the corresponding gradient filters $\frac{\partial d_{x:ij}}{\partial x}$ and $\frac{\partial d_{y:ij}}{\partial y}$, respectively.  To shorten the time required for the dose optimization, preceding the dose optimization, we prepare the filter tables of dose gradients, $-\frac{(x - x_j)}{\sqrt{2\pi}\sigma_x^3}\exp\left(-\frac{(x - x_j)^2}{2\sigma_x^2}\right)$ and $-\frac{(y - y_j)}{\sqrt{2\pi}\sigma_y^3}\exp\left(-\frac{(y - y_j)^2}{2\sigma_y^2}\right)$, as a function of standard deviations $\sigma_x$ and $\sigma_y$, in addition to those of the normalized Gaussians, $\frac{1}{\sqrt{2\pi}\sigma_x}\exp\left(-\frac{(x - x_j)^2}{2\sigma_x^2}\right)$ and $\frac{1}{\sqrt{2\pi}\sigma_y}\exp\left(-\frac{(y - y_j)^2}{2\sigma_y^2}\right)$.

As examples, the normalized Gaussian functions with standard deviations of 3.0, 4.0 and 5.0 mm are shown in figure 1 along with the corresponding functions of their derivatives.  These filter tables are referenced in the dose optimizations with the indices of $\sigma_x$ and $\sigma_y$ determined for a given pencil beam $j$ at a given depth $z$.

Similarly, preceding the dose optimization, the values of $\partial\sigma_{x:j}(z_i)/\partial z$, $\partial\sigma_{y:j}(z_i)/\partial z$ and $\dfrac{\partial d_{z:ij}}{\partial z}$ for each pencil beam $j$ in equation (12) are calculated in addition to the widths $\sigma_{x:j}(z_i)$ and $\sigma_{y:j}(z_i)$ and the dose $d_{z:j}(z_i)$. The derivatives of $\sigma_{x:j}(z)$ and $\sigma_{y:j}(z)$ with respect to $z$, $\partial\sigma_{x:j}(z)/\partial z$ and $\partial\sigma_{y:j}(z)/\partial z$, can easily be obtained as

$$\frac{\partial\sigma_{x:j}(z)}{\partial z} \approx \frac{\sigma_{x:j}((z+\Delta z))-\sigma_{x:j}(z)}{\Delta z} \qquad (13)$$

and

$$\frac{\partial\sigma_{y:j}(z)}{\partial z} \approx \frac{\sigma_{y:j}((z+\Delta z))-\sigma_{y:j}(z)}{\Delta z}. \qquad (14)$$

As examples, the lateral beam widths $\sigma_{x:j}(z)$ and $\sigma_{y:j}(z)$ of a 290-MeV/u carbon beam and corresponding derivatives $\partial\sigma_{x:j}(z)/\partial z$ and $\partial\sigma_{y:j}(z)/\partial z$ are shown in figure 2. The derivative $\dfrac{\partial d_{z:ij}}{\partial z}$ can be derived from the dose distribution $d_{z:j}(z)$ as

$$\frac{\partial d_{z:j}(z)}{\partial z} \approx \frac{d_{z:j}((z+\Delta z))-d_{z:j}(z)}{\Delta z}, \qquad (15)$$

as shown in figure 3 for a 290-MeV/u carbon beam. These precalculated values are used in the dose optimization to shorten the time required to derive the robust dual-field plan.

4. Simulations

4.1 Treatment plan

To demonstrate the effectiveness of the proposed algorithm, we made treatment plans for three target volumes located in an oval-shaped phantom (240 mm along the major axis and 200 mm along the minor axis with 100-mm height) as shown in figure 4. The phantom is assumed to be homogeneous and water equivalent. The voxel size was $\Delta x=\Delta y=\Delta z=2.0$ mm. We consider different dual-field geometries for respective target volumes: a unidirectional geometry for target-1, an opposing geometry for target-2 and an orthogonal geometry for target-3. For dose calculation and dose delivery, we assumed a 290-MeV/u carbon beam. The maximum range of the scanned carbon beam is 151.6 mm in water. The range of the beam is shifted using range shifter plates. The effective field-size is set to be 150 mm square in the transverse directions. Spots are placed on a regular, rectangular grid, with 4-mm spacing in both the beam- and the transversal-directions for each field.

The optimizations were done with and without applying the proposed method for each of three targets where the parameters $D_P^{max}$, $D_P^{min}$, $Q_P^o$ and $Q_P^u$ were fixed to 2.0 Gy, 2.0 Gy, 1.0 and 1.0, respectively. In the following discussion, the former plan is referred to as a 'robust plan', while the latter plan is referred to as a 'conventional plan'. In the robust plan, we tentatively set the penalty coefficients $Q_x$, $Q_y$ and $Q_z$ to 50.0, 50.0 and 100.0, respectively. Qualitatively speaking, these penalties should be defined to be proportional to the uncertainties in the respective directions.

IMPT is strongly dependent on the choice of initial beam weights (starting conditions) used for the optimization (Lomax 1999, Oelfke and Bortfeld 2000, Albertini et al 2007). We predetermine the initial beam weights so as to deliver a flat spread-out Bragg peak (SOBP) type profile of 1 Gy from each beam direction using the algorithm described in Krämer et al (2000).

4.2 Recalculation of the treatment plan

For the range and setup uncertainties analysis, the resultant six treatment plans, i.e., the robust and the conventional plans for each of the three targets described in section 4.1, were recalculated with combinational geometrical perturbations. The range was intentionally misplaced by varying the phantom effective density by ±5%. Similarly, for setup uncertainties, the dose distribution of each of two fields was calculated with intentional translations of ±5 mm in both the *x*- and *y*-directions. Including the nominal range and position, $3^5$ = 243 possible combinations of the total dose distribution were derived for each plan.

5. Results and Discussion

5.1 Unidirectional geometry

Figure 5 shows treatment plans optimized for target-1 with a unidirectional-field geometry. The upper row (figure 5(1)) shows the conventional plan, and the lower row (figure 5(2)) shows the robust plan. Figures 5(1b) and 5(2b) are the resultant dose distributions when the nominal range and setup position are realized for each field as assumed for the optimization. In both plans, the planned dose distributions with a flatness of ±3% were realized in the target volume encompassed with a yellow solid line. In the robust plan, a 'safety margin' is created at the distal, proximal and lateral field edges by the method described in section 3.1. The dose contributions of the individual beams are shown in figure 5(1a, 1c) and 5(2a, 2c). In the conventional plan, there are two distinct regions of high dose gradient within the individual fields. Their locations correspond to the width limits of the beam scanning. Owing to the predetermined initial beam weights, the dose within the overlap region was kept flat at 1 Gy in each field. On the other hand, in the robust plan, a gradual dose distribution was observed at the overlap region in each field, as shown in figures 5(2a) and (2c), with evenly spaced isodose lines.

To show the effectiveness of our method for reducing the sensitivity of the treatment plans to range and setup uncertainties in a unidirectional-field geometry, we recalculated 243 possible combinations of perturbed dose distribution for each of the two plans. The maximum and minimum dose distributions, defined as $D_{max:i} = \underset{k=1-243}{\text{Max}}\{D_{i,k}\}$ and $D_{min:i} = \underset{k=1-243}{\text{Min}}\{D_{i,k}\}$ (Lomax 2008), were derived for both plans, and shown in figures 6(1a, 2a) and 6(1b, 2b), respectively. These two distributions can be considered to provide the positive and negative 'error bars' of possible doses on either side of the nominal plans shown in figure 5(1b) and 5(2b). In the conventional plan, hot doses up to 2.72 Gy and cold doses down to 1.28 Gy appeared around the regions where distinct internal dose gradients were observed (see figures 5(1a) and 5(1c)). To show the dose reduction clearly, we displayed the deviation between the minimum dose distribution and the prescribed one within the target volume in figures 6(1c) and 6(2c) for the conventional and robust plans, respectively. A significant dose reduction of 1.87 Gy was observed in the distal and lateral field edges in the conventional plan, in addition to the cold doses caused by the steep dose gradients at the field junctions, as shown in figure 6(1c). Over- and under-dosages were also observed in the robust plan within the overlap region of the two fields (figures 6(2a) and 6(2b)). However, because of the gradual dose distribution of each field, they were up to 2.28 Gy and down to 1.72 Gy, respectively. Furthermore, the cold dose at the peripheral region of the target was significantly reduced because of the expansion of the target region in step 1, as shown in figure 6(2c).

The dose volume histograms (DVHs) of 243 dose distributions are plotted in figure 7 for both plans. Figure 7 also shows clearly the reduced sensitivity to range and setup uncertainties of the robust plan compared to the conventional plan. The 95% doses (D95) were 86.0% (1.72 Gy) and 46.5% (0.93 Gy) for the robust and conventional plans, respectively, in the worst cases.

As described above, in the conventional plan, hot and cold doses were observed at the regions where distinct in-field dose gradients exist. However, the predetermined initial Bragg peak weights provide similar dose distributions to those of the multiple-patch technique used in the passive irradiation method (Hug et al 2000) to smear the undesired hot and cold doses near the patch-field junction. Hence, even the conventional plan described in this study provides rather a robust plan compared to the normal field-patching technique.

5.2 Opposing geometry

Figure 8 shows treatment plans optimized for target-2 with an opposing-field geometry. The upper and lower rows of the figure show the conventional and robust plans, respectively. Figures 8(1b) and 8(2b) are the resultant dose distributions when the nominal ranges and setup positions are realized for each field as assumed for the optimization. The planned dose distributions were realized in the target volume for both plans. In the robust plan, the high dose region was expanded into the outer region of the target to ensure the target coverage for range and setup errors. The dose contributions of the individual beams are shown in figures 8(1a, 1c) and 8(2a, 2c), respectively. In this irradiation, the maximum penetration depths of the FTVs are 180 mm in water equivalent length (mmWEL), which is slightly larger than the range of a 290-MeV/u carbon beam. In the conventional plan, two distinct regions of high dose gradient were observed within the individual fields, corresponding to the maximum ranges of the opposing beams. On the other hand, in the robust plan, a gradual dose distribution was produced at the overlap region in each field, as shown in figures 8(2a) and (2c) with evenly spaced isodose lines.

The maximum and minimum dose distributions were derived for the conventional and robust plans for target-2 with an opposing-field geometry as shown in figures 9(1a, 2a) and 9(1b, 2b), respectively. In the conventional plan, hot doses up to 2.56 Gy and cold doses down to 1.47 Gy were observed in the regions where distinct high dose gradients exist (see figures 8(1a) and 8(1c)). Furthermore, in the conventional plan, a dose reduction of 1.64 Gy was observed in the peripheral region of the target, as shown in figure 9(1c), showing the deviations between the minimum dose distribution and the prescribed one within the target volume. On the other hand, in the robust plan, over- and under-dosages were observed only in the overlap region of the two fields, as shown in figures 9(2a) and 9(2b). Owing to the gradual in-field dose distribution, the undesired hot and cold doses observed in the conventional plan were smeared in the robust plan. The hot and cold doses were up to 2.29 Gy and down to 1.71 Gy, respectively.

The DVHs of 243 dose distributions are plotted in figure 10 for the two plans. D95 in the worst case was increased from 60.5% (1.21 Gy) to 83.8% (1.68 Gy) by applying the proposed robust optimization method.

5.3 Orthogonal geometry

Figure 11 shows treatment plans optimized for target-3 with an orthogonal-field geometry. The upper and lower rows of the figure show the conventional and robust plans, respectively. Figures 11(1b) and 11(2b) are the resultant nominal dose distributions showing good target coverage for both plans. A 'safety margin' was created at the distal, proximal and lateral field edges in the robust plan by the method described in section 3.1. The dose distributions of individual fields are displayed in figures 11(1a, 1c) and 11(2a, 2c), respectively. In the robust plan, the dose profile in the overlap region resembles the shape of a spiral staircase, making the dose gradient as shallow as possible.

The maximum and minimum dose distributions were derived for the conventional and the robust plan as shown in figures 12(1a, 2a) and 12(1b, 2b), respectively. The dose delivered to the lower left part of the target in the axial image becomes relatively insensitive to range and setup variations, especially in the robust plan. However, in both plans, hot and cold doses were possible at the inner side of the field junction. In the conventional plan, a hot dose up to 2.81 Gy and a cold dose down to 1.14 Gy were found around that region, while they are 2.66 Gy and 1.35 Gy in robust plan, respectively. These findings are similar to those described in Unkelbach et al (2007), where the probabilistic robust algorithm has been applied to optimize the dose distribution for the dual-field geometry using an RTOG benchmark phantom. In the conventional plan, contrary to the robust plan, a dose reduction of 1.80 Gy was also found in the peripheral region of the target, as shown in figure 12(1c), illustrating the deviations between the minimum dose distribution and the prescribed one within the target volume.

The DVHs of 243 dose distributions are plotted in figure 13 for the two plans. D95 in the worst cases were 59.8% (1.21 Gy) for the conventional plan and 76.2% (1.68 Gy) for the robust plan.

5.4 Computation time

Optimization of the conventional dual-field plans took 7, 6 and 3 minutes (on a Dell Precision 690 workstation with 3.0 GB RAM) for the unidirectional, opposing and orthogonal geometries, respectively. On the other hand, optimization of the robust plans took 24, 22 and 10 minutes for these geometries. The total number of spots, $N_{spots}$, is 23290 (33936), 17816 (24624) and 9792 (15036) in the conventional (robust) plan, and the total number of voxels within the DTV, $N_{T}$, is 115659 (161464), 77529 (121393) and 39709 (68335) for these geometries, respectively. The size of the problem defined by $N_{spots} \times N_{T}$ in the robust plan is approximately twice that of the problem in the conventional plan. Consequently, roughly speaking, the expansion of the target volume (Step 1) prolongs the dose convolution time by a factor of two. Further prolongation of the computation time was caused by the introduction of the gradient suppression terms in the objective function (Step 2).

5.5 Outlook

For simplicity, we applied the proposed robust algorithm to the optimization of the physical dose distribution in the dual-field geometry with a scanned carbon beam. However, this algorithm can be applied to physical dose optimization in three or more fields geometry with any heavy charged-particle beam including a proton beam. Our algorithm can also be applied to treatment planning for a photon IMRT.   To shorten the time required for the dose optimization, we prepared filter tables of the lateral dose profile of the scanned pencil beam and its gradient preceding the dose optimizations.   In this study, the lateral dose profile of the scanned pencil beam was expressed as a Gaussian function and its gradient was analytically derived.   However, this method can also be used for a non-Gaussian beam model in which the gradient cannot be derived analytically, by expressing the gradient numerically.

A robust approach for IMPT/IMIT including the critical structures will be described in our next paper.   The assessment of the proposed method in a biological dose optimization scenario remains a future study.

6.   Conclusions

The quality of the dual-field strategy is quite sensitive to range and setup uncertainties. In this paper, we described an algorithm to reduce the sensitivity of the dual-field plan to these uncertainties in scanning irradiations.   The algorithm is composed of two steps: 1) generating the expanded target volume, and 2) solving the inverse problem where the terms suppressing the dose gradient of individual fields are added into the objective function.   The former step reduces the undesired cold dose observed in the peripheral region of the target, while the latter step mitigates the significant hot and cold doses generated at the field junctions.   The effectiveness of the algorithm was demonstrated for three extreme cases of dual fields with unidirectional, opposing and orthogonal geometries.   The proposed method greatly reduced the sensitivity of the dual-field plan to range and setup errors.   The 95% doses were increased to 86.0%, 83.8% and 76.2% from 46.5%, 60.5% and 59.8% for these geometries with the robust algorithm, respectively.   Compared to the conventional dual-field plan, we observed an increase of the optimization time for the proposed robust plan by a factor of approximately three.

References

Albertini F, Bolsi A, Lomax A J, Rutz H P, Timmermann B and Goitein G 2008 Sensitivity of intensity modulated proton therapy plans to changes in patient weight *Radiat. Oncol.* **86** 187-194

Bussiere M R and Adams J A 2003 Treatment planning for conformal proton radiation therapy *Technol. Cancer Res. Treat.* **2** 389-399

Furukawa T, Inaniwa T, Sato S, Shirai T, Takei Y, Takeshita E, Mizushima K, Iwata Y, Himukai T, Mori S, Fukuda S, Minohara S, Takada E, Murakami T and Noda K 2010 Performance of NIRS fast scanning system for heavy-ion radiotherapy *Med. Phys.* submitted

Furukawa T, Inaniwa T, Sato S, Shirai T, Mori S, Takeshita E, Mizushima K, Himukai T and Noda K 2010 Study of moving target irradiation with fast rescanning and gating in particle therapy *Med. Phys.* submitted

Hug E B, Adams J, Fitzek M, De Vries A and Munzenrides J E *et al* 2000 Fractionated, three-dimensional, planning-assisted proton-radiation therapy for orbital rhabdomyosarcoma: a novel technique *Int. J. Radiat. Oncol. Biol. Phys.* **47** 979-984

Inaniwa T, Furukawa T, Sato S, Tomitani T, Kobayashi M, Minohara S, Noda K and Kanai T 2008 Development of treatment planning for scanning irradiation at HIMAC *Nucl. Instrum. Methods. Phys. Res. B* **266** 2194-2198

Inaniwa T, Furukawa T, Nagano A, Sato S, Saotome N, Noda K and Kanai T 2009 Field-size effect of physical doses in carbon-ion scanning using range shifter plates *Med. Phys.* **36** 2889-2897

Inaniwa T, Furukawa T, Tomitani T, Sato S, Noda K and Kanai T 2007 Optimization for fast-scanning irradiation in particle therapy *Med. Phys.* **34** 3302-3311

Kanai T *et al* 1999 Biophysical characteristics of HIMAC clinical irradiation system for heavy-ion radiation therapy *Int. J. Radiat. Oncol. Biol. Phys.* **44** 201-210

Kase Y, Kanai T, Matsumoto Y, Furusawa Y, Okamoto H, Asaba T, Sakama M and Shinoda H 2006 Microdosimetric measurements and estimation of human cell survival for heavy-ion beams *Radiat. Res.* **166** 629-638

Krämer M and Scholz M 2000 Treatment planning for heavy-ion radiotherapy: calculation and optimization of biologically effective dose *Phys. Med. Biol.* **45** 3319-3330

Li Y, Zhang X, Dong Lei and Mohan R 2007 A novel patch-field design using an optimized grid filter for passively scattered proton beams *Phys. Med. Biol.* **52** N265-ZN275

Lomax A J 1999 Intensity modulation methods for proton radiotherapy *Phys. Med. Biol.* **44** 185-205

Lomax A J, Boehringer T, Coray A, Egger E, Goitein G, Grossmann M, Juelke P, Lin S, Pedroni E, Rohrer B, Roser W, Rossi B, Siegenthaler, Stadelmann O, Stauble H, Vetter C and Wisser L 2001 Intensity modulated proton therapy: a clinical example *Med. Phys.* **28** 317-324

Lomax A J 2008 Intensity modulated proton therapy and its sensitivity to treatment uncertainties 2:

the potential effects of inter-fraction and inter-field motions *Phys. Med. Biol.* **53** 1043-1056

Oelfke U and Bortfeld T 2000 Intensity modulated radiotherapy with charged particle beams: studies of inverse treatment planning for rotation therapy *Med. Phys.* **27** 1246-1257

Pflugfelder D, Wilkens J J and Oelfke U 2008 Worst case optimization: a method to account for uncertainties in the optimization of intensity modulated proton therapy *Phys. Med. Biol.* **53** 1689-1700

Paganetti H, Jiang H, Parodi K, Slopsema R and Engelsman M 2008 Clinical implementation of full Monte Carlo dose calculation in proton beam therapy *Phys. Med. Biol.* **53** 4825-4853

Rutz H P, Weber D C, Goitein G, Ares C, Bolsi A, Lomax A J, Pedroni E, Coray A, Hug E B and Timmermann B 2008 Prostate spot-scanning proton radiation therapy for chordoma and chondrosarcoma in children and adolescents: initial experience at Paul Scherrer Institute *Int. J. Radiat. Oncol. Biol. Phys.* **71** 220-225

Unkelbach J, Chan T C Y and Bortfeld T 2007 Accounting for range uncertainties in the optimization of intensity modulated proton therapy *Phys. Med. Biol.* **52** 2755-2773

Unkelbach J, Bortfeld T, Martin B C and Soukup M 2009 Reducing the sensitivity of IMPT treatment plans to setup errors and range uncertainties via probabilistic treatment planning *Med. Phys.* **36** 149-163

Weber D C, Rutz H P, Pedroni E S, Bolsi A, Timmermann B, Verwey J, Lomax A J and Goitein G 2005 Results of spot-scanning proton radiation therapy for chordoma and chondrosarcoma of the skull base: the Paul Scherrer Institute experience *Int. J. Radiat. Oncol. Biol. Phys.* **63** 401-409

Figure Captions

Figure 1. One-dimensional Gauss/differential Gauss filter used in treatment planning software. (a) One-dimensional Gauss filter with standard deviations of 3 mm (solid-line), 4 mm (dashed-line) and 5 mm (dotted-line), respectively. (b) One-dimensional differential Gauss filter with standard deviations of 3 mm (solid-line), 4 mm (dashed-line) and 5 mm (dotted-line), respectively.

Figure 2. (a) The lateral beam widths of a 290-MeV/u carbon beam in the *x*-direction (solid line), $\sigma_x$, and the *y*-direction (dotted line), $\sigma_y$, as a function of depth $z$ for range shifter thicknesses of 0 (black), 30 (red), 60 (green) and 90 (blue) mm water equivalent thickness, respectively. (b) The corresponding derivatives of the lateral beam widths $\partial \sigma_x(z)/\partial z$ and $\partial \sigma_y(z)/\partial z$.

Figure 3. (a) The planner-integrated doses of a 290-MeV/u carbon beam $d_z(z)$ as a function of depth $z$ for range shifter thicknesses of 0 (black), 30 (red), 60 (green) and 90 (blue) mm water equivalent thickness, respectively. (b) The corresponding derivatives of planner-integrated doses with respect to *z*, $\partial d_z(z)/\partial z$.

Figure 4. Geometries of three target volumes (dark gray) located in the oval-shaped phantom (light gray). The thick arrows indicate the beam directions delivered in each target.

Figure 5. The conventional treatment plan (upper row) and the robust treatment plan (lower row) optimized for target-1 with a unidirectional-field geometry. The resultant dose distributions are shown for the two plans with a color wash display in (1b) and (2b), where the yellow line outlines the target. The dose distributions delivered by individual beams are displayed in (1a, c) and (2a, c) with isodose lines, where the target volume is identified with dark yellow.

Figure 6. Maximum dose distributions (1a, 2a) and minimum dose distributions (1b, 2b) for the conventional and robust plans, respectively. The yellow line outlines the target in (1a), (2a), (1b) and (2b). The deviations between the minimum dose distribution and the prescribed one within the target volume are shown in (1c) and (2c) for the two plans.

Figure 7. DVH of dose distributions recalculated for 243 different realizations of the beam ranges and field positions for the conventional (black curves) and robust plans (red curves), respectively.

Figure 8. The conventional treatment plan (upper row) and the robust treatment plan (lower row) optimized for target-2 with an opposing-field geometry. The resultant dose distributions are shown for the two plans with a color wash display in (1b) and (2b), where the yellow line outlines the target. The dose distributions delivered by individual beams are displayed in (1a, c) and (2a, c) with isodose lines, where the target volume is identified with dark yellow.

Figure 9. Maximum dose distributions (1a, 2a) and minimum dose distributions (1b, 2b) for the conventional and robust plans, respectively. The yellow line outlines the target in (1a), (2a), (1b) and (2b). The deviations between the minimum dose distribution and the prescribed one within the target volume are shown in (1c) and (2c) for the two plans.

Figure 10. DVH of dose distributions recalculated for 243 different realizations of the beam ranges and field positions for the conventional (black curves) and robust plans (red curves), respectively.

Figure 11. The conventional treatment plan (upper row) and the robust treatment plan (lower row) optimized for target-3 with an orthogonal-field geometry. The resultant dose distributions are shown for the two plans with a color wash display in (1b) and (2b) where the yellow line outlines the target. The dose distributions delivered by individual beams are displayed in (1a, c) and (2a, c) with isodose lines, where the target volume is identified with dark yellow.

Figure 12. Maximum dose distributions (1a, 2a) and minimum dose distributions (1b, 2b) for the conventional and robust plans, respectively. The yellow line outlines the target in (1a), (2a), (1b) and (2b). The deviations between the minimum dose distribution and the prescribed one within the target volume are shown in (1c) and (2c) for the two plans.

Figure 13. DVH of dose distributions recalculated for 243 different realizations of the beam ranges and field positions for the conventional (black curves) and robust plans (red curves), respectively.

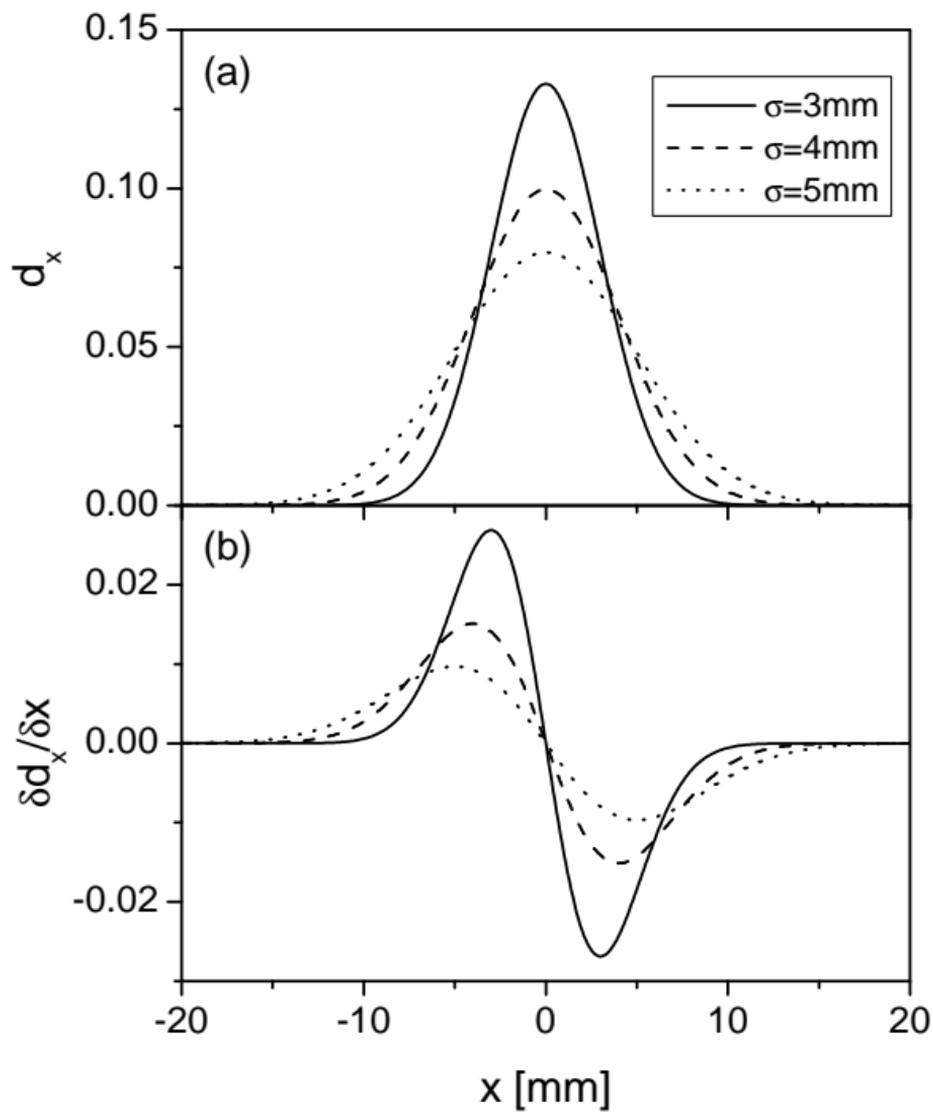

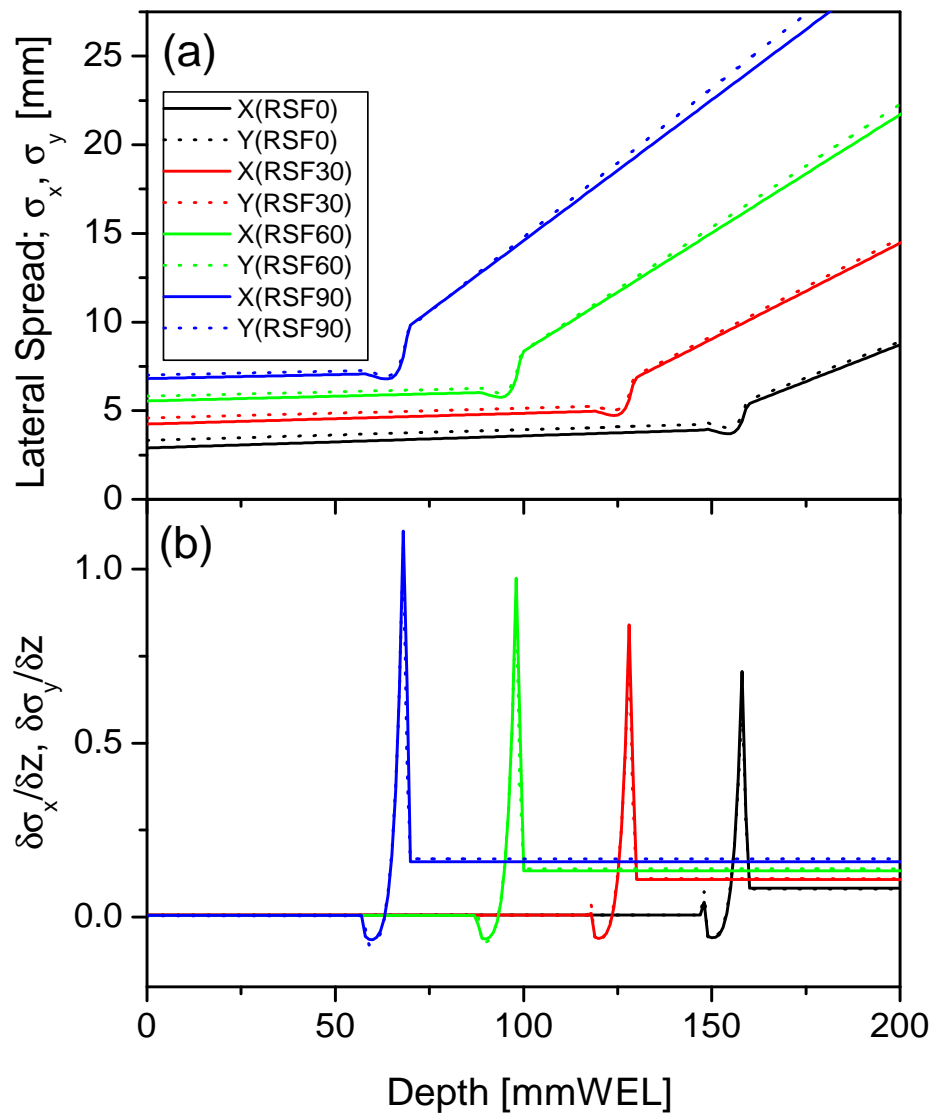

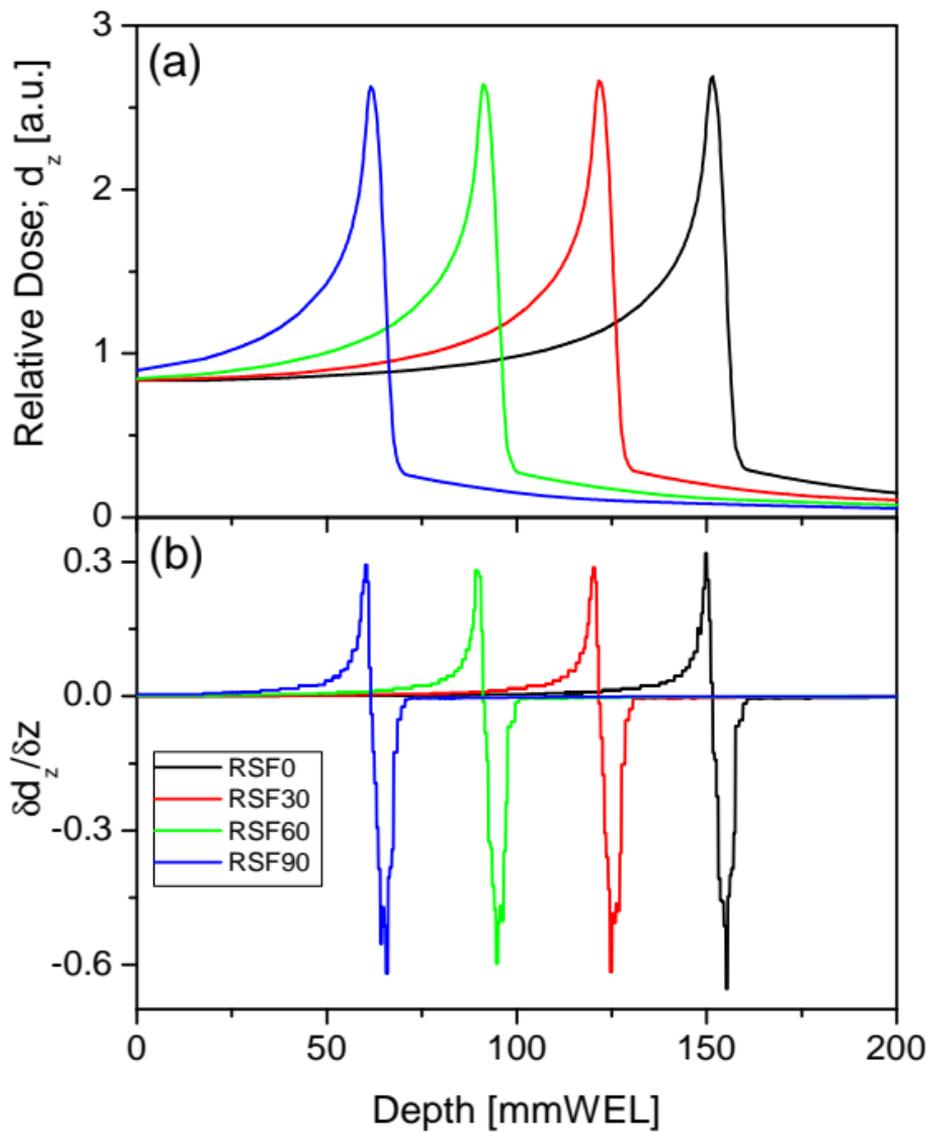

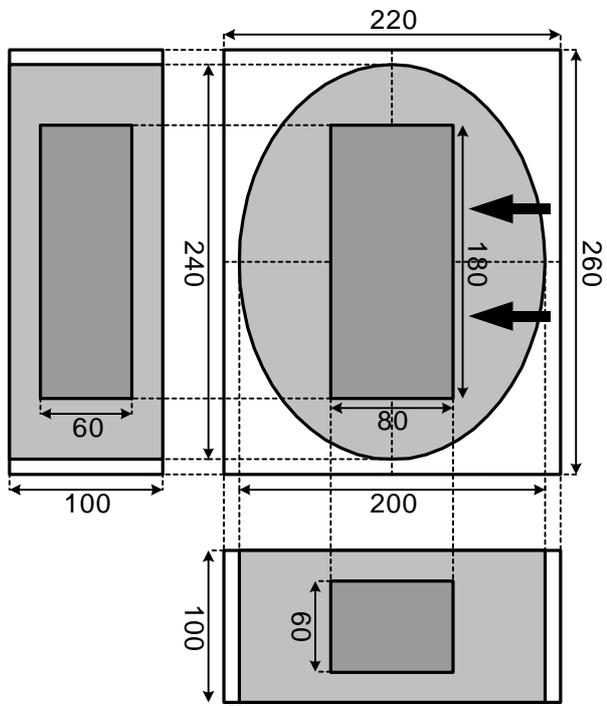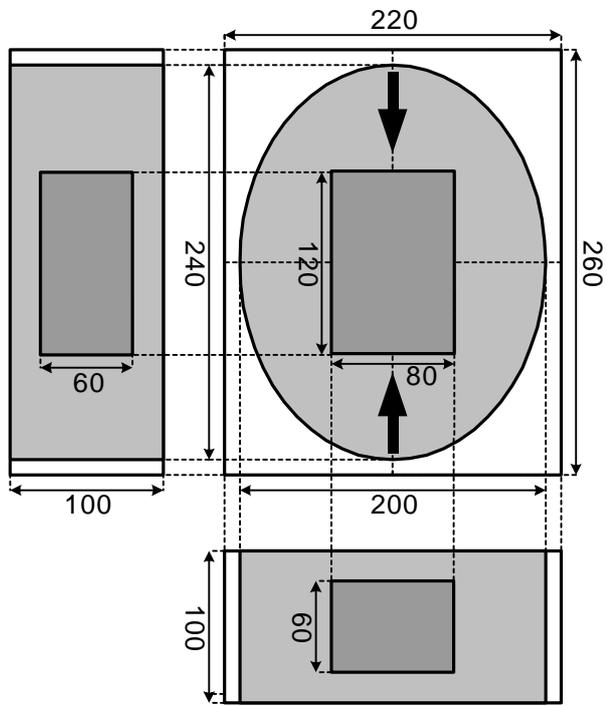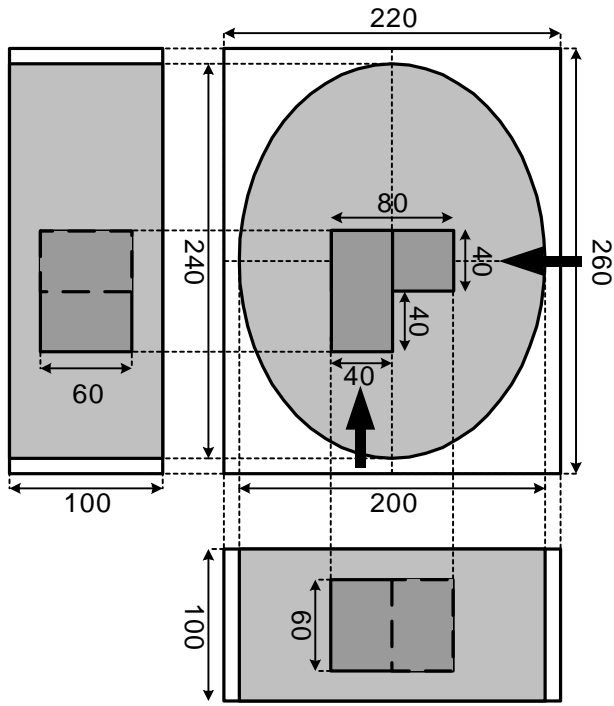

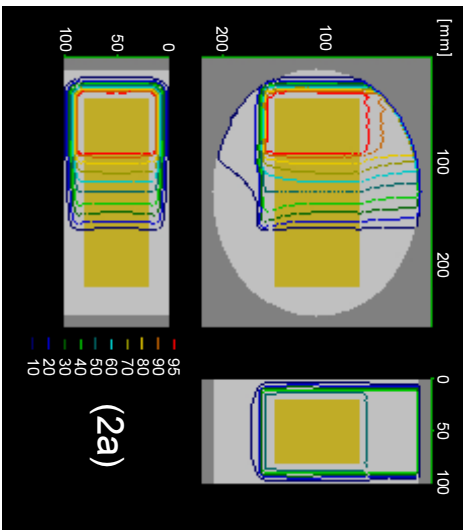
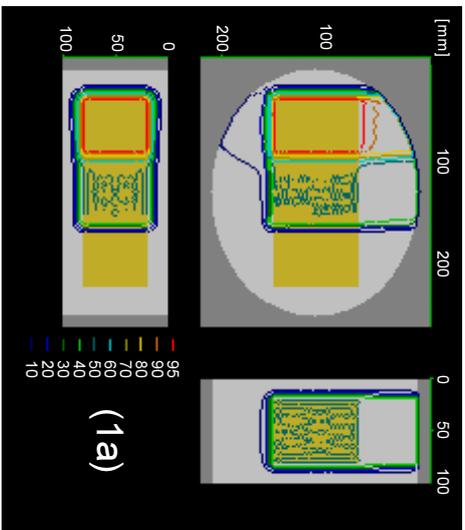

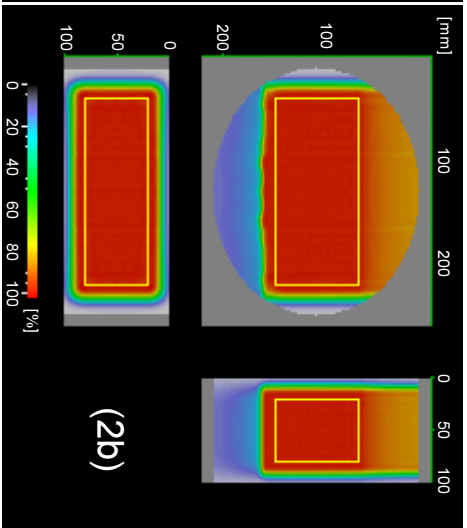
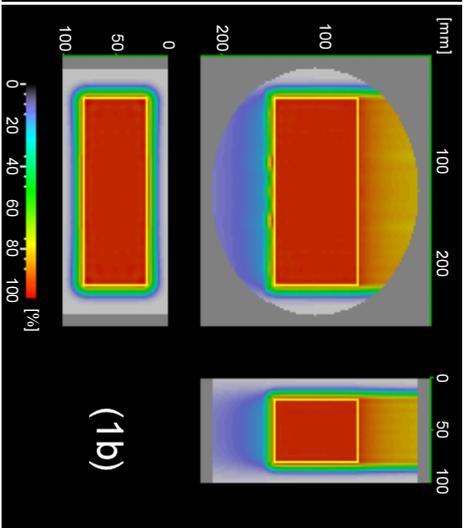

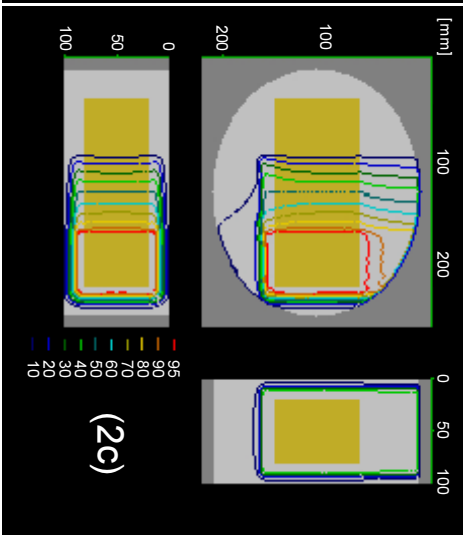
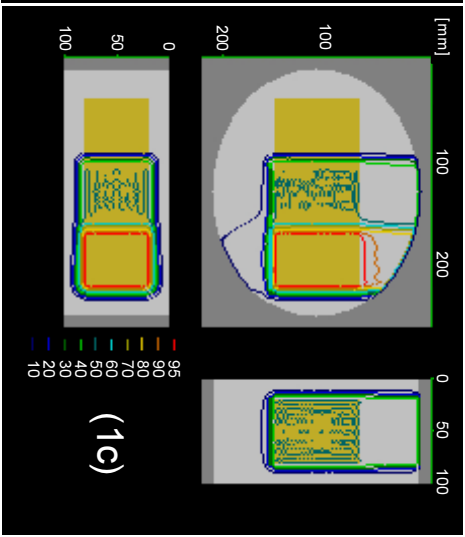

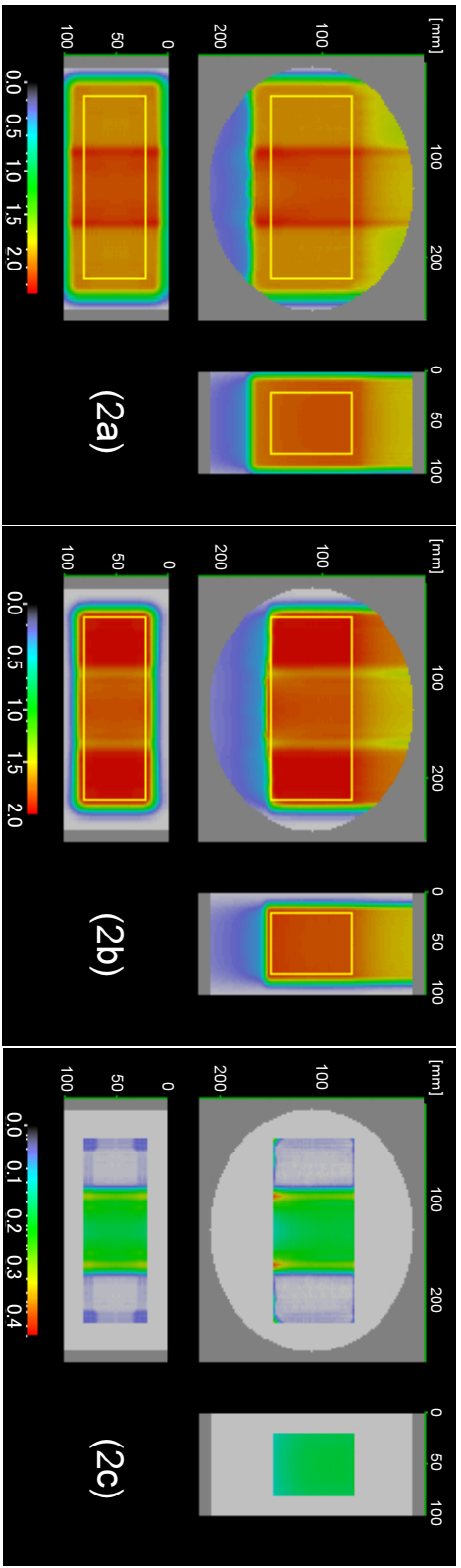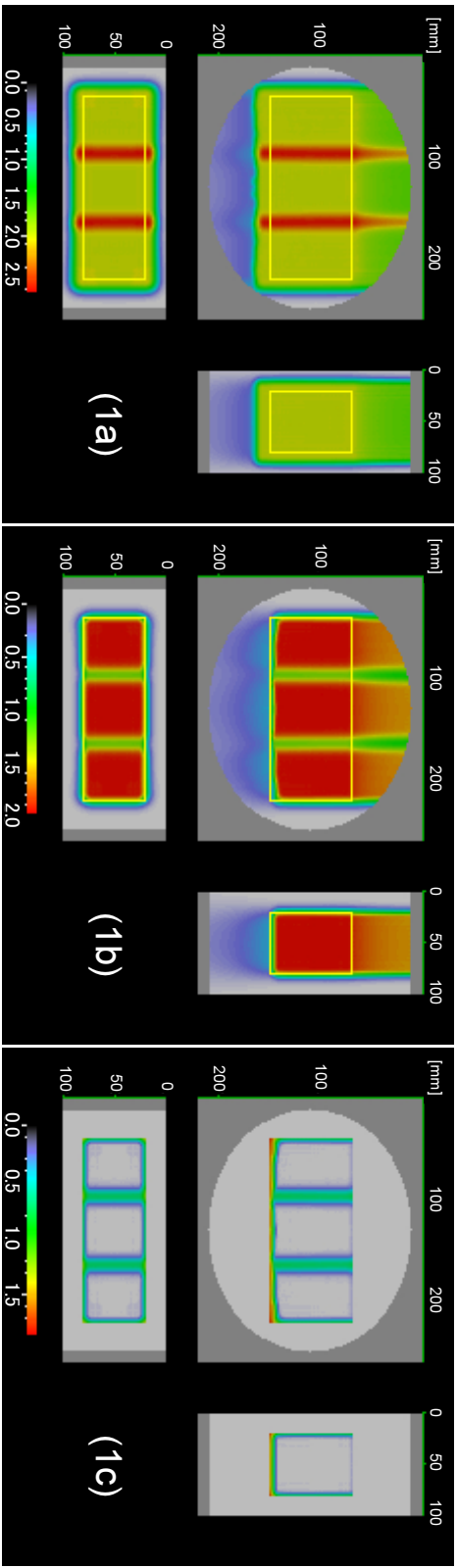

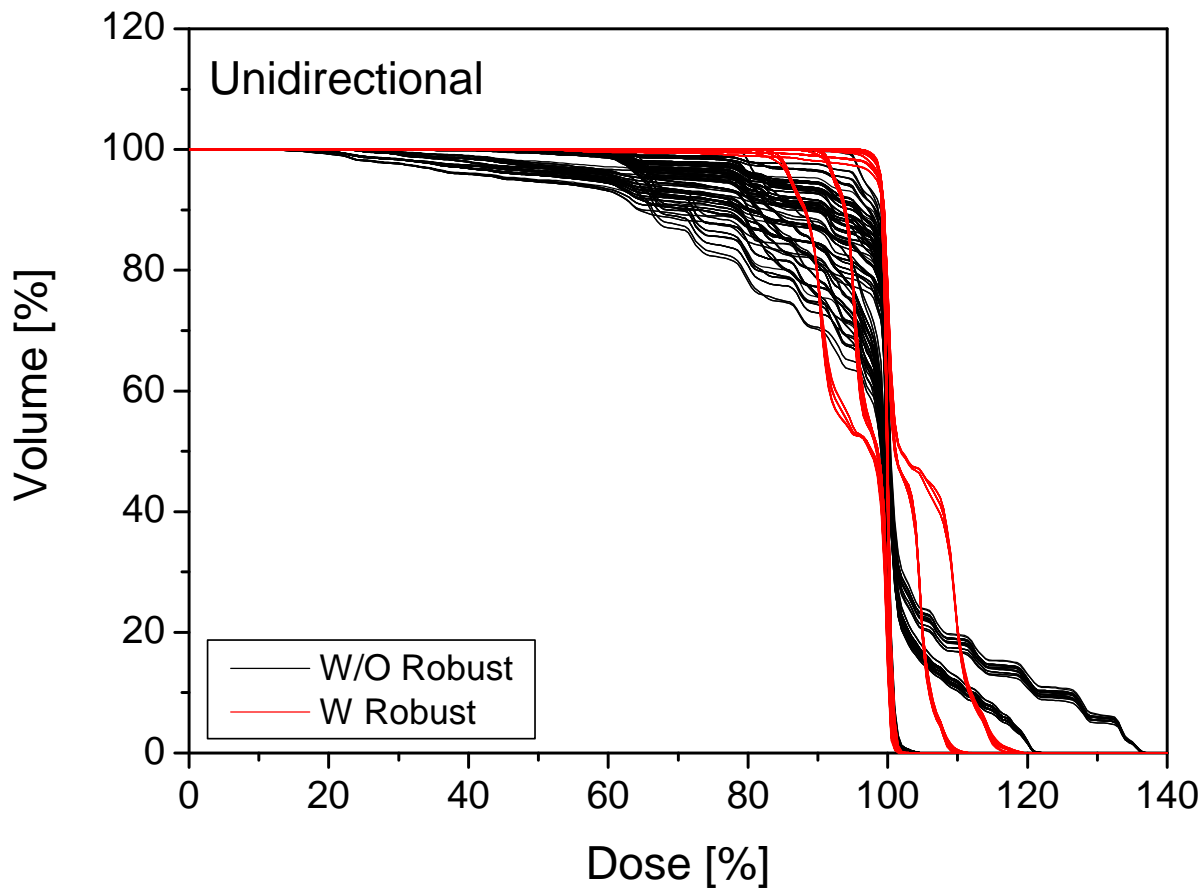

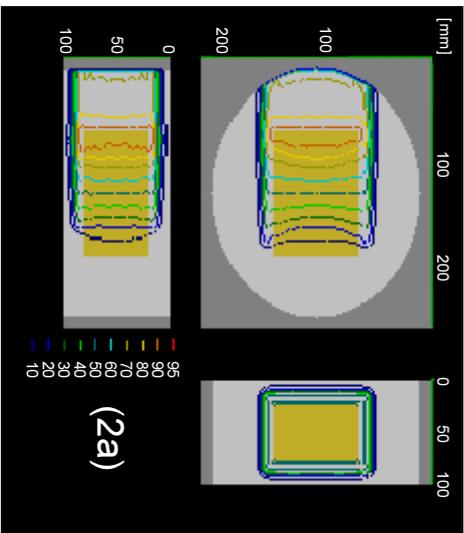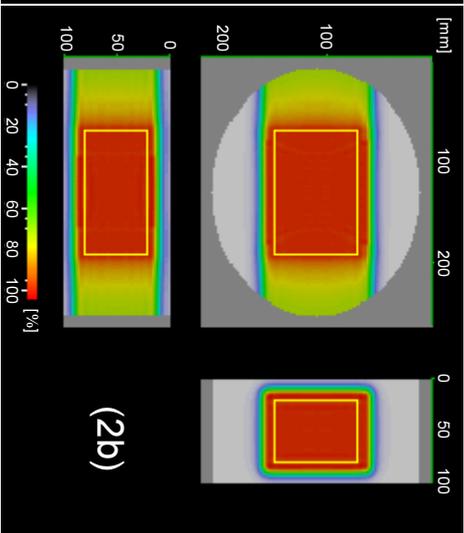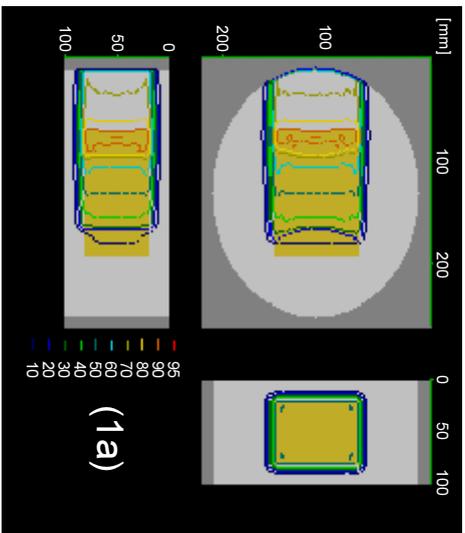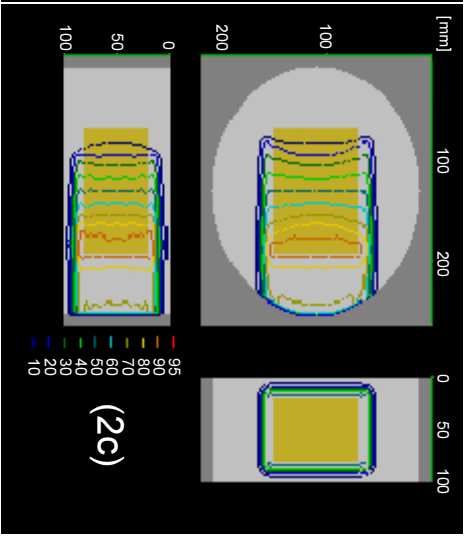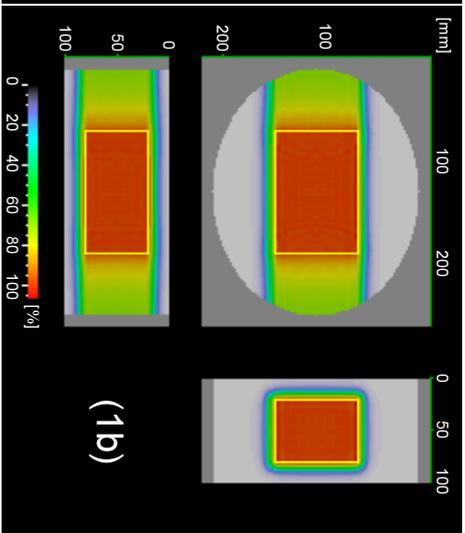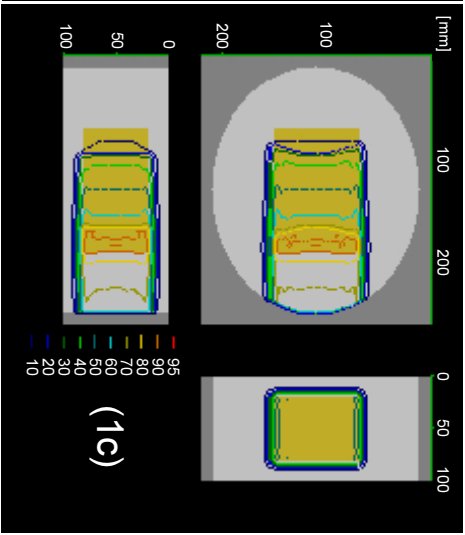

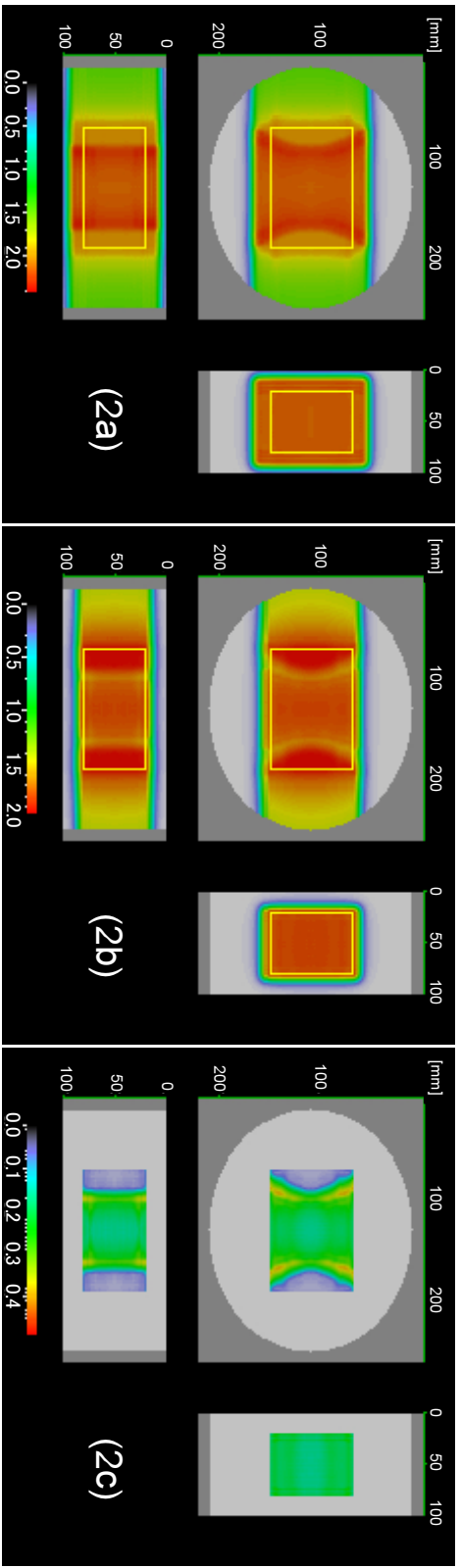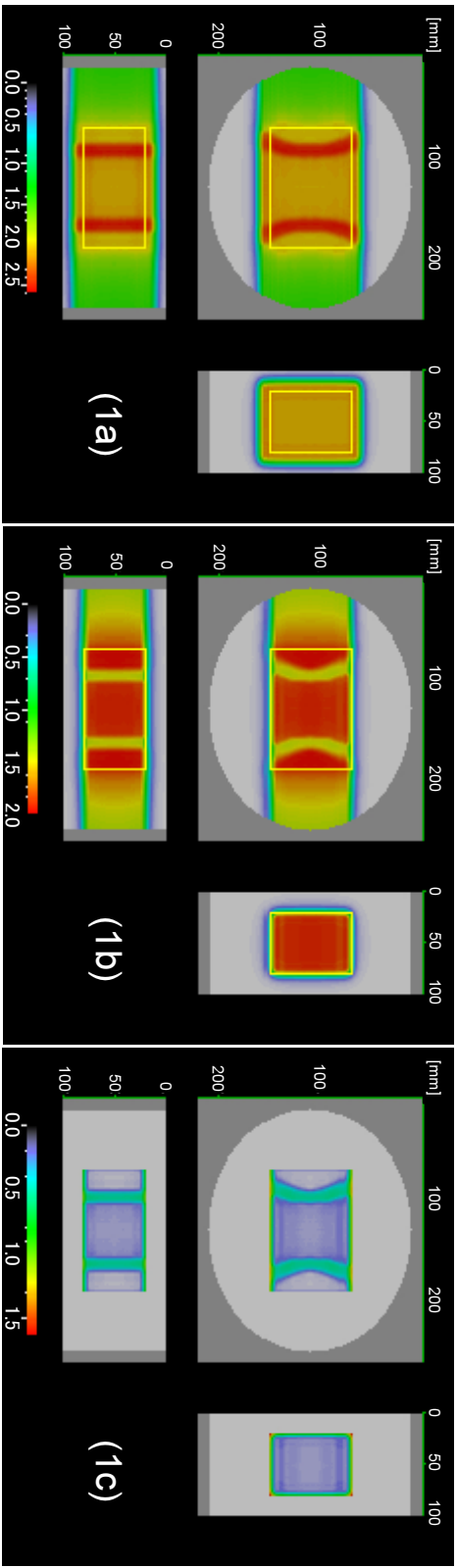

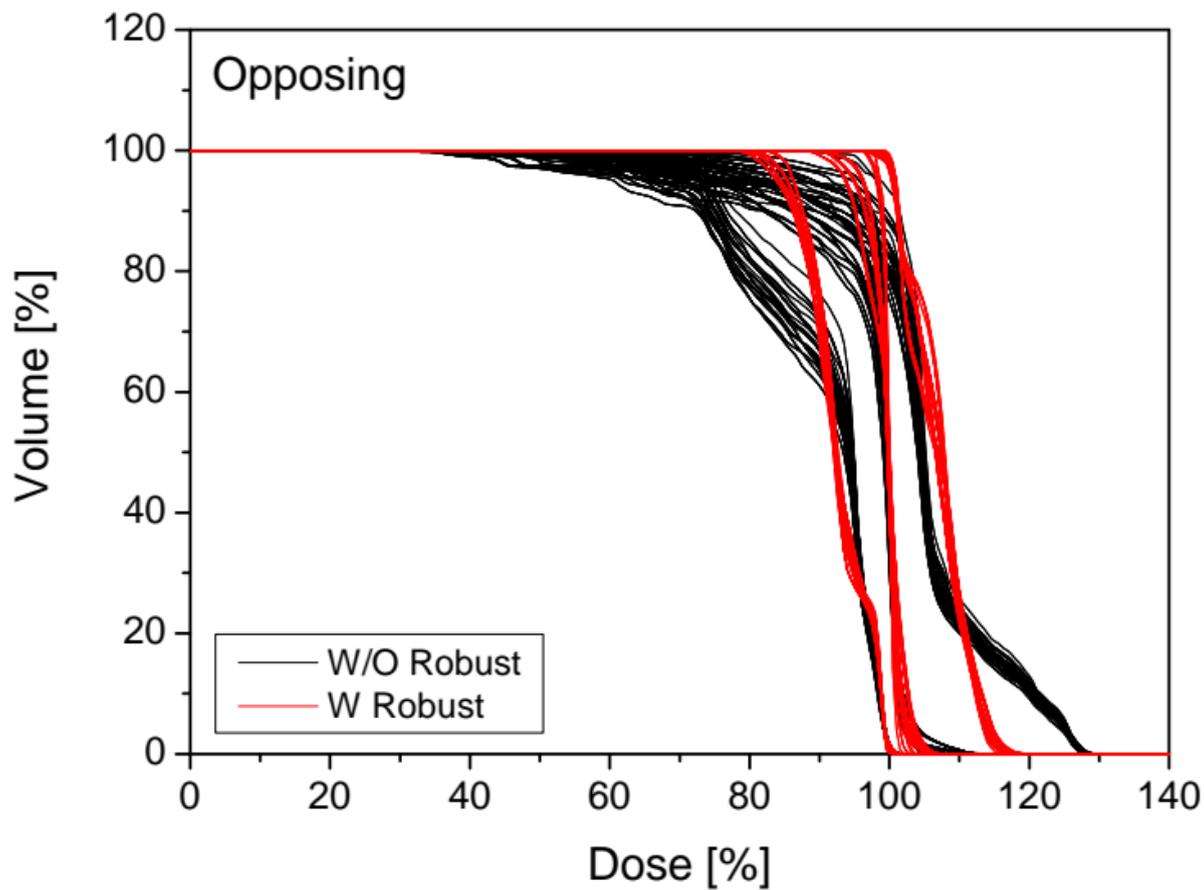

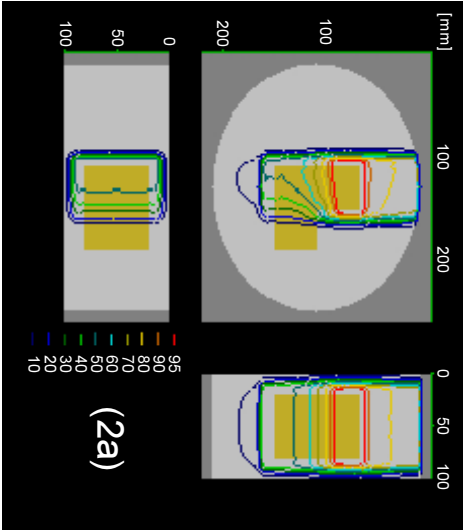 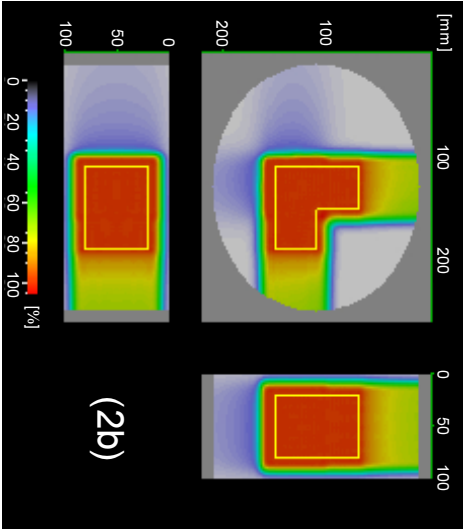 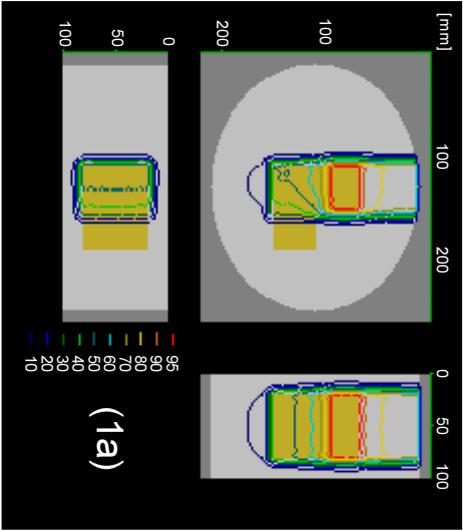

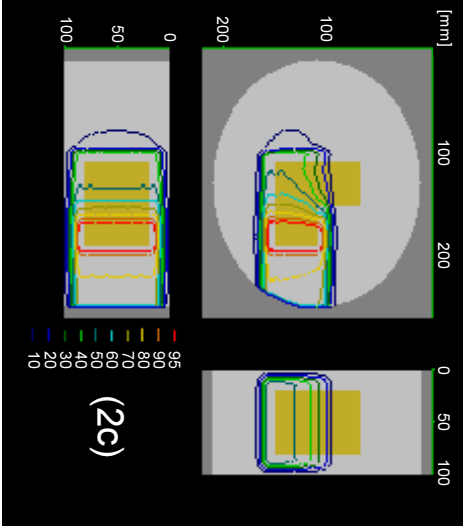 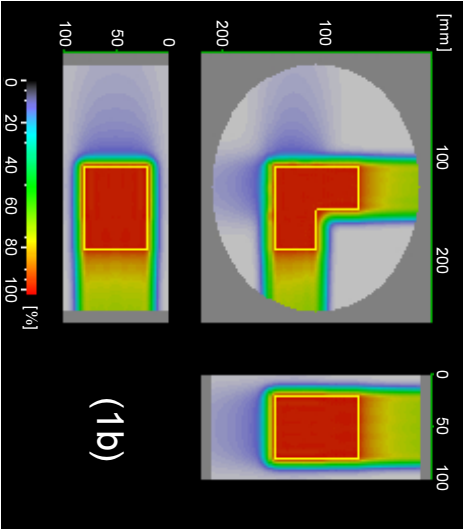 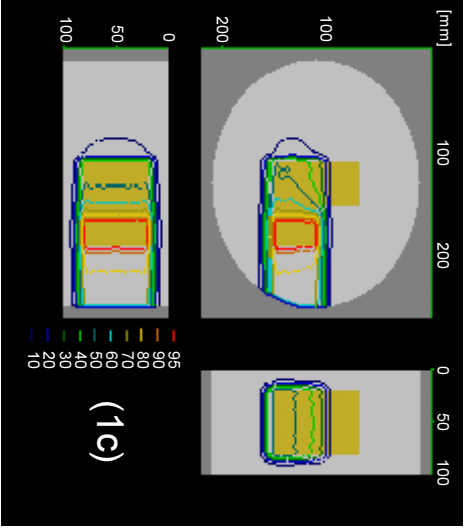

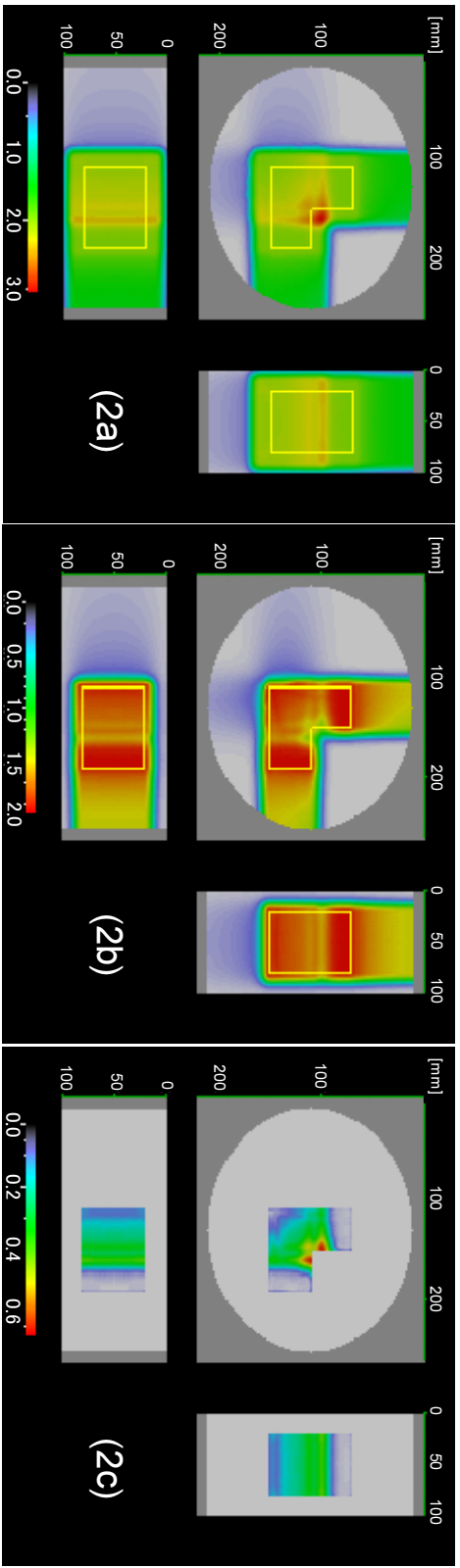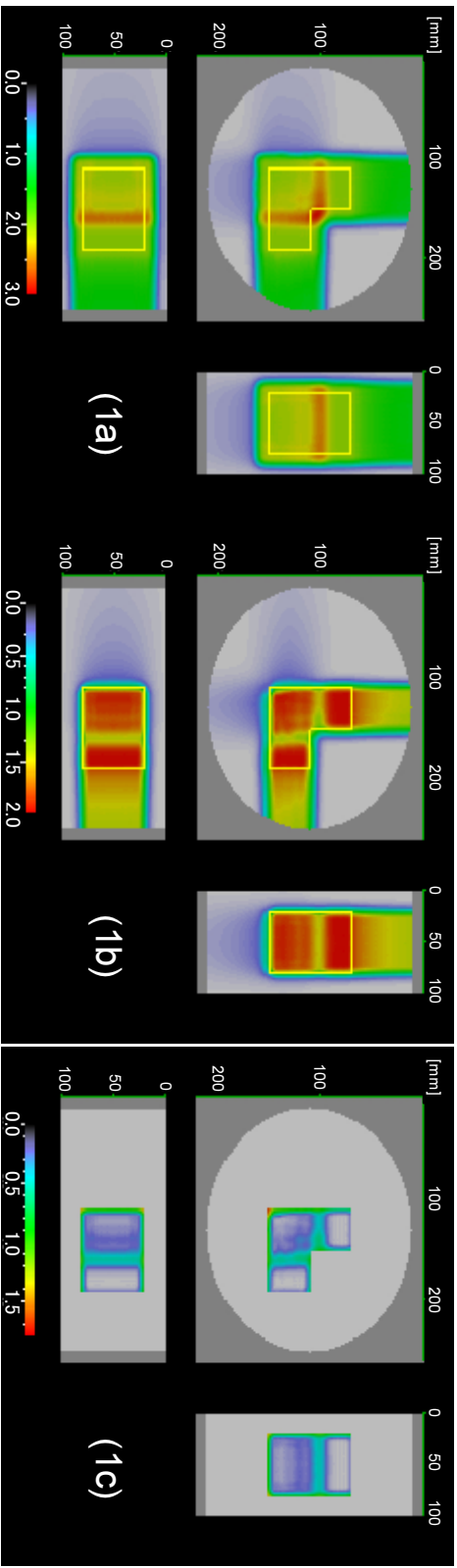

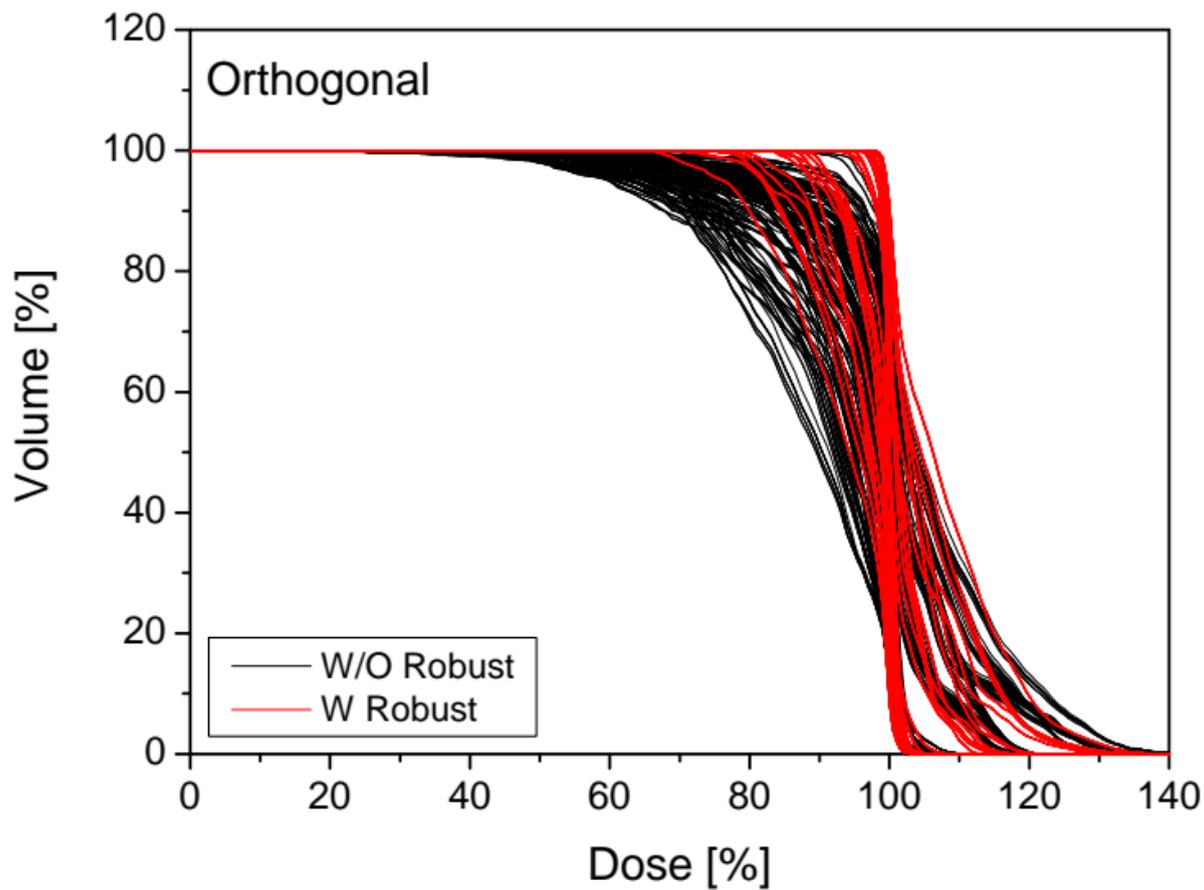